\documentclass[12pt,runningheads]{llncs}
\usepackage[utf8]{inputenc}
\usepackage{amsmath, amsfonts}
\usepackage{fullpage}
\usepackage{natbib}
\usepackage{pifont}
\usepackage{color}
\usepackage{graphicx}
\usepackage{hyperref}
\newcommand{\cmark}{\ding{51}}%
\newcommand{\xmark}{\ding{55}}%
\newcommand{\cost}{\mbox{cost}}
\title{\textbf{Optimal sampling and assay for soil organic carbon estimation}}
\author{Jacob V. Spertus}
\institute{
University of California, Berkeley, Department of Statistics\\
\email{jakespertus@berkeley.edu}
}
\date{November 2020}

\begin{document}

\maketitle

\begin{abstract}
    The world needs around 150 Pg of negative carbon emissions to mitigate climate change. Global soils may provide a stable, sizeable reservoir to help achieve this goal by sequestering atmospheric carbon dioxide as soil organic carbon (SOC). In turn, SOC can support healthy soils and provide a multitude of ecosystem benefits. To support SOC sequestration, researchers and policy makers must be able to precisely measure the amount of SOC in a given plot of land. SOC measurement is typically accomplished by taking soil cores selected at random from the plot under study, mixing (compositing) some of them together, and analyzing (assaying) the composited samples in a laboratory. Compositing reduces assay costs, which can be substantial. Taking samples is also costly. Given uncertainties and costs in both sampling and assay along with a desired estimation precision, there is an optimal composite size that will minimize the budget required to achieve that precision. Conversely, given a fixed budget, there is a composite size that minimizes uncertainty. In this paper, we describe and formalize sampling and assay for SOC and derive the optima for three commonly used assay methods: dry combustion in an elemental analyzer, loss-on-ignition, and mid-infrared spectroscopy. We demonstrate the utility of this approach using data from a soil survey conducted in California. We give recommendations for practice and provide software to implement our framework.
\end{abstract}

\section{Introduction}

Climate change is likely to put enormous strain on nature and human societies in the coming decades. It is largely driven by the release of atmospheric carbon dioxide (CO$_2$) that was once sequestered in the earth, either in fossil fuels or as soil carbon. Since the cultivation of soils began, soils have lost about 50-70\% of their carbon to the atmosphere. Soil still accounts for the 2nd largest store of carbon on Earth after the ocean, containing about 7.5 times that of the atmosphere \citep{lal_soil_2018}. However, agriculture is now one of the largest contributors to global carbon emissions.       

In pursuit of solutions, a growing movement of farmers and other advocates are highlighting ``regenerative agriculture" as a way to make agriculture a net sink, rather than a source, of carbon -- drawing CO$_2$ from the atmosphere and sequestering it in the land as soil organic carbon (SOC). Regenerative agriculture provides a variety of ecosystem services including water use efficiency, biodiversity, and overall soil health. These may be sufficient to support its use, but to pay for regenerative agriculture on the basis of SOC sequestration, decision makers need to know \textit{how much} carbon is sequestered by different strategies.

In order to measure SOC sequestration, at a minimum scientists must be able to measure how much SOC is in a given plot of land at a given point in time. This task is referred to as ``SOC stock estimation." Soil scientists accomplish SOC stock estimation by collecting multiple cores of soil from a given plot, preparing/processing the samples, and analyzing (assaying) their SOC concentration by a number of different techniques. SOC is either presented on its own, as a concentration, or it is converted to stock using soil bulk density measured on nearby intact cores. Both sampling and assay of SOC concentration are subject to uncertainties and both become expensive at the volumes necessary to overcome these uncertainties. All else equal, increasing the number of samples and the number of assays will reduce uncertainty while driving up costs. A process called compositing allows investigators to reduce cost by mixing together sampled cores and assaying the mixture(s), but incurs additional error when there is uncertainty in the assay. 

\begin{figure}[h]
    \centering
    \includegraphics[width = \textwidth]{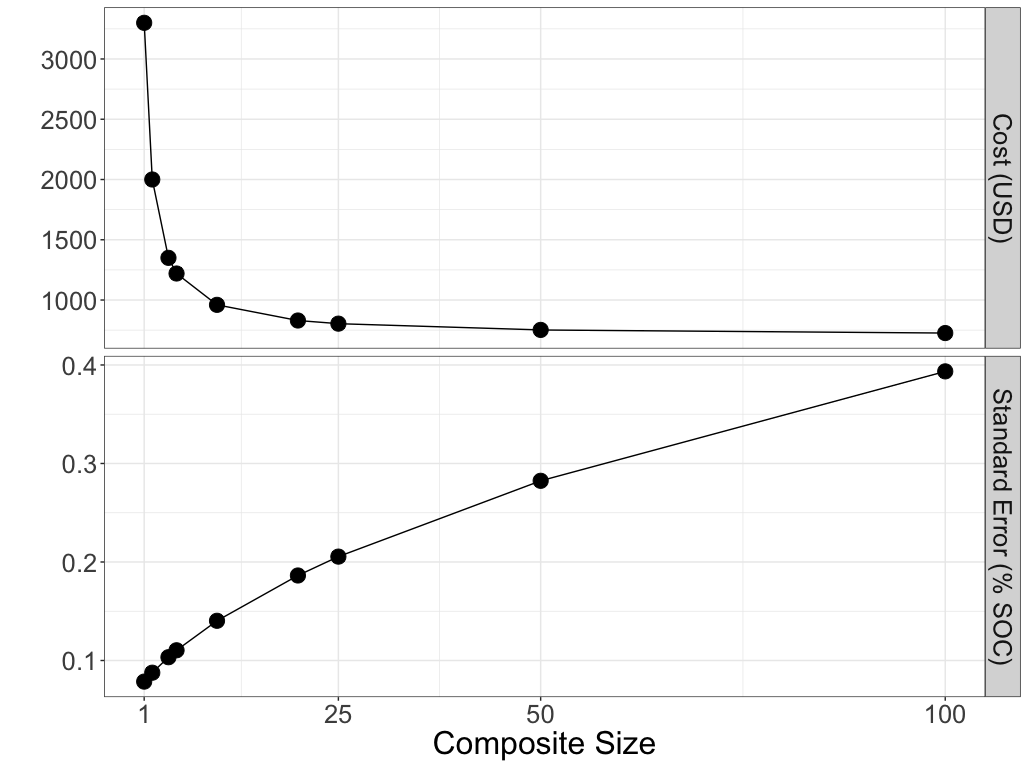}
    \caption{Standard error and costs associated with estimating SOC concentration across a range of possible composite sizes. Taking more assays decreases the standard error but increases the overall cost of estimation. Inputs are given in table \ref{tab:inputs}. Specifically, we display compositing trade-offs for assaying topsoil SOC concentration (true average is 3.57\% SOC) by loss-on-ignition (see Sections \ref{sec:assay} and \ref{sec:application}). SOC = soil organic carbon; USD = United States Dollars.}
    \label{fig:composite_sizes}
\end{figure}

Figure \ref{fig:composite_sizes} sketches this trade-off in an example where 100 cores have been collected, and the investigator must now choose how much to composite before assaying the composited samples. Parameters and costs are taken from a survey of California rangelands, detailed later in this paper (Section \ref{sec:application}). Across the range of possible composite sizes, the standard error of the average SOC concentration estimate decreases by a factor of 5, while the cost increases by a factor of 7. Clearly, compositing has substantial implications for both uncertainty and cost. 

In this paper we resolve this trade-off by presenting sampling and assay as an optimization problem. Given a fixed budget, we derive the sampling and assay sizes that minimize estimation uncertainty. Conversely, given a fixed estimation precision we'd like to achieve, we derive the optimal sizes to minimize the budget. The solutions depend on the heterogeneity and mean SOC concentration of the plot(s) under study, the assay error, and the costs associated with sampling and assay.

 Our paper is organized as follows. In Section \ref{sec:lit_review} we situate our work in the soil science and statistics literature. In Section \ref{sec:estimation_goal} we formalize the objectives of SOC estimation. We then turn to the logistics and statistics of estimation, covering sampling in Section \ref{sec:sampling}, compositing in Section \ref{sec:compositing}, sample preparation in Section \ref{sec:sample_prep}, and assay in Section \ref{sec:assay}. Section \ref{sec:optimization} contains our main results. In Section \ref{sec:application} we apply our results to data, demonstrating their use in practice. Section \ref{sec:discussion} discusses additional nuances, challenges, and extensions of stock estimation, and provides recommendation for practice. All of our work is supported by R software, available at \url{https://github.com/spertus/soil-carbon-simulations}. An online graphical user interface is available at \url{https://jakespertus.shinyapps.io/shiny/}.

\section{Literature Review}
\label{sec:lit_review}

As part of this paper we review the components of stock estimation and the processes of sampling, compositing, and assay. We focus on estimating the average concentration of SOC in a plot. In order to make minimal assumptions about the plot under study and for our results to be as general as possible, we take the design-based perspective on estimation. Thus, the model of SOC concentration in the plot is minimal. Specifically, we do not make any assumptions about the spatial distribution of SOC concentration. Inference proceeds from random sampling, while SOC concentration is unknown but fixed. \citet{webster_field_2012} and \citet{gruijter_sampling_2006} provide accessible reviews of soil sampling, inference, and optimization from the design-based perspective. 

The design-based perspective contrasts with the model-based or ``geostatistical" perspective, originally developed to map gold mines  \citep{krige_statistical_1951}. The geostatistical approach to SOC stock estimation conceptualizes SOC content as random or at least well approximated by a random process. Geostatistics is especially useful for estimating an entire function of a soil property, i.e. for mapping. We do not examine the model-based approach in detail here. \citet{diggle_model-based_2007} and \citet{gruijter_sampling_2006} are good references on geostatistics and its applications to natural resource monitoring.

\citet{patil_composite_2011} provides a detailed accounting of the statistics of compositing, and includes an analysis of compositing with additive assay error. The benefits of compositing depend on the relative size of the plot heterogeneity to the assay error. \citet{lark_considerations_2012} analyzes properties of various compositing schemes alongside a geostatistical model for spatial variation. The author shows that compositing nearby cores improves the precision of an SOC map, compared to taking a single core at each location. \citet{kosmelj_comparison_2001} analyzes compositing alongside a cost model in the context of soil sampling for zinc or calcium, solving an optimization problem for compositing over subplots without considering assay error. In a case study, they found that optimal compositing could reduce costs by around 50\% while maintaining estimation precision.


We analyze three laboratory assay methods used to measure SOC concentration in soil samples: loss-on-ignition (LOI), dry combustion in an elemental analyzer (DC-EA), and mid-infrared spectroscopy (MIRS). LOI involves measuring the difference in mass before and after heating samples in a furnace. The heating cooks off the organic matter in the soil -- along with an unpredictable amount of ``mineral" or structural water. The amount of mass lost can be mapped to the SOC concentration in the sample using ordinary least squares regression \citep{nayak_current_2019, vos_capability_2005}. DC-EA combusts small aliquots of soil at high temperatures in an elemental analyzer that measures the amount of CO$_2$ released during the burn. DC-EA machines vary in their specifics, but are generally considered the gold-standard for precise determination of SOC concentration \citep{nayak_current_2019, fao_measuring_2020, smith_how_2020}. MIRS assays carbon by shining infrared light on samples and recording the wavelengths absorbed. These wavelengths (``spectra") can then be closely mapped to SOC concentration (determined by DC-EA) using machine learning methods. MIRS requires a considerable upfront investment both in the machinery and in developing a large spectral library that links wavelength signatures to SOC concentrations within a region of interest (e.g. a country or state). MIRS and vis-NIRS could become highly cost-effective assay strategies as prices come down and spectral libraries expand \citep{england_proximal_2018, nayak_current_2019, wijewardane_predicting_2018}. LOI and MIRS are ``high-throughput" methods, as many samples can be analyzed quickly and cheaply. However these methods offer less precision than DC-EA, and may be prone to biases.

The core contribution of this paper is similar in spirit to a classical power analysis, which determines how many samples are needed to estimate quantities to within a desired precision or to run a hypothesis test at a desired power. \citet{kravchenko_whole-profile_2011} presents basic methods and an application of power analysis to detecting SOC change in tillage experiments. \citet{pringle_soil_2011} derived sample sizes necessary to detect changes in SOC stocks on Australian rangelands. A 2019 report by the Food and Agriculture Organization of the United Nations also includes a section on conducting power analysis \citep{fao_measuring_2020}. These power analyses do not consider the effects of compositing or assay error, nor do they consider the costs of sampling and assay. In our work we provide a framework to derive optimal composite sizes given a cost model. In the process, we characterize budgets that are needed to achieve reasonable precision when estimating SOC concentration.

There is a precedent for analyzing optimal designs in soil science, but most of this work has been done in the geostatistical literature and generally concerns how to optimally distribute samples given an assumed model. If scientists have access to a reliable variogram describing the spatial distribution of SOC, then the sampling design can be optimized to minimize estimation or prediction variance \citet{van_groenigen_constrained_1999, brus_chapter_2006}. If SOC exhibits \textit{any} spatial auto-correlation, well-spread random samples can increase efficiency compared to uniform independent random sampling. Traditionally, grid or transect sampling is often used, but these designs may be biased and don't yield accurate standard errors \citet{webster_field_2012, wolter_investigation_1984}. Investigators may also use auxiliary variables, like management type, topography, or vegetation, to yield more efficient sampling designs. \citet{de_gruijter_farm-scale_2016} presents a recipe to estimate SOC concentration or stock at the farm scale. That paper focuses on reducing costs through an optimally-stratified sampling design, while compositing receives less attention. Other modern design approaches aim to improve spatial coverage or auxiliary variable balance through sophisticated random sampling. Well-spread random samples can be achieved by a kind of nested stratification, as in the generalized random tessellation stratified design \citep{stevens_spatially_2004}, or by the cube or local pivotal method, wherein samples repel each other spatially \citep{tille_probability_2016}. All of these papers seek to optimally distribute sample points and do not account for assay error.

New ways of measuring SOC stocks continue to emerge at a rapid pace, driven by advances in technology and data science. Assay can now be accomplished directly in the field using techniques like mobile infrared spectroscopy, eddy covariance assay, inelastic neutron scattering, and laser-induced breakdown spectroscopy. These techniques tend to involve far more assay error than laboratory analyses \citep{chatterjee_evaluation_2009,nayak_current_2019,england_proximal_2018}. Additionally, an active area of research seeks to combine various assays and remote sensor data using machine learning and geostatistics \citep{wadoux_multi-source_2019, padarian_using_2019, england_proximal_2018}. A few of these new technologies do not involve randomly sampling cores, and are thus outside the context of this work. The rest apply readily to framework we present here.

\section{What are we trying to estimate?}
\label{sec:estimation_goal}
SOC concentration (e.g. percent SOC or grams of SOC per kg of soil) is a (non-random) three-dimensional function in latitude, longitude, and depth. In this paper, we are interested in estimating the average concentration, $\mu$, in a bounded area of land to some fixed depth; or the total stock of SOC $\mathcal{T}$ in the area. Typically, estimation occurs within fixed depth profiles, which can then be aggregated to whole-profile stock or concentration estimates. The equivalent soil mass method provides an important alternative strategy wherein profiles are defined to some predetermined mass, not depth  \citep{wendt_equivalent_2013}. 

We follow the convention of estimating concentrations and stocks \textit{within} profiles defined by depth or mass. We thus suppress dependence on depth as we develop our ideas. For concreteness, the reader may imagine we are only discussing top-soil concentration or stock in what follows, though our analysis applies to any profile.   
We also stress that the maximum depth of the survey is very important.
Many physical, chemical, and biological mechanisms can move SOC downward or cause soil loss at depth. Long-term management can impact deep soil SOC, so concentrations and stocks may need to be estimated down to a meter or more to accurately account for the SOC sequestration of different management strategies \citep{tautges_deep_2019, luo_can_2010}. 

If we are only interested in average concentration, it suffices to estimate $\mu$.
If we want to estimate the stock $\mathcal{T}$, we also need the bulk density in grams per cubic centimeter $d$, the area of the plot in square meters $\mathcal{A}$, and the length of the profile in meters $L$. Assuming that bulk density is constant within depth, the total amount of carbon within the depth profile is 

$$\mathcal{T} \equiv 10^4 \times L \times \mathcal{A} \times \mu \times d.$$

The factor 10$^4$ includes conversion of \%SOC to gram per gram, and bulk density to grams per cubic meter. Different factors may be applied to report SOC in tons per hectare (Mg ha$^{-1}$).

In reality, SOC is never exactly the same across a study area. The degree of heterogeneity can be expressed as the plot variance, $\sigma_p^2$, which is the average squared distance of SOC concentration from the mean $\mu$ (for a definition in symbols see Section \ref{sec:framework} in the Appendix). If every point in the plot has the same SOC concentration $\mu$, then $\sigma_p = 0$. On the other hand, if the SOC concentration is highly variable across the plot then $\sigma_p$ will be large. The maximum value, $\sigma_p = 50$, is attained when half the plot is 0\% SOC and the other half is 100\% SOC. Along with the assay precision, the plot heterogeneity $\sigma_p$ allows us to characterize the uncertainty in estimates of $\mu$.

\section{Sampling}
\label{sec:sampling}

Investigators typically estimate $\mu$ by sampling relatively small amounts of soil from the plot under study.
Soil samples can be taken using an auger, a corer, or by digging a pit. An augers can mix soil horizons, while with a corer horizons are typically kept distinct. Compaction can occur with either method, which may skew depths or density estimates. Digging a pit and drawing samples from the side may yield the best samples, with clear horizons and no compaction, but is relatively destructive and very labor intensive. In what follows, we typically refer to a distinct (uncomposited) sample as a ``core," though in principle it could be drawn by any of the above methods.  
 Taking cores at randomly sampled locations can ensure that estimates of $\mu$ are unbiased. In this section, we describe three random sampling approaches that are regularly used in practice: uniform independent random sampling, stratified sampling, and cluster sampling.

Uniform independent random samples (UIRSs) are generated by sampling $n$ points uniformly --- no particular locations are favored --- and independently --- the location of a particular core does not affect the location of any other. In the soil science literature, UIRSs are sometimes equated with ``simple random samples" \cite{webster_field_2012}. However, in statistics simple random sampling denotes uniform sampling without replacement from a discrete, finite population. We use the more cumbersome UIRS to avoid confusion. Sometimes, plots are conceptually ``discretized" by mapping the continuous surface to a fine grid, which then becomes the finite sampling frame so that simple random sampling is equivalent to uniform independent random sampling (UIRSing). UIRSs can provide unbiased estimates of $\mu$ no matter how SOC is distributed in the plot. UIRSs also yield unbiased estimates of the heterogeneity $\sigma_p$. This allows researchers to characterize the precision of the estimate and thus to conduct hypothesis tests or construct confidence intervals based on a UIRS.  

Stratified sampling can be used to take advantage of auxiliary information about the distribution of SOC, which can yield more precise estimates. For example, in rangeland the distribution of SOC may be driven by topography, vegetation type, mineralogy, microclimates, or land-use history \citep{pringle_soil_2011, webster_field_2012}. Strata and sample sizes per strata can be selected using algorithms that predict SOC concentrations in order to maximize the expected precision given a fixed overall sample size \citep{de_gruijter_farm-scale_2016}. Like UIRSs, stratified samples can yield unbiased estimates of $\mu$, $\sigma_p$, and the variance of estimators. 

Finally, cluster random samples are drawn by first choosing a point at random and then deterministically sampling along a regular transect or grid extending from the original point. Cluster random samples with a single random starting point are sometimes called ``systematic random samples" in the soil science literature \citep{gruijter_sampling_2006}. Cluster random samples have the advantage of automatically distributing samples evenly across part of a plot. Logistically, this makes samples relatively easy to collect, since cores can be efficiently taken by moving regular distances along the transect or grid. Statistically, this reduces the variance of sample means from cluster random samples when SOC is positively correlated in space, a standard geostatistical assumption. However, sample means from cluster random samples are not inherently unbiased and do not have a simple variance. Both of these properties depend on further assumptions about how SOC is distributed within the plot \citep{gruijter_sampling_2006, webster_field_2012, wolter_investigation_1984}. If these assumptions are not met, cluster random samples may yield biased or imprecise estimates. Periodicity of the property under study (due to row cropping, for example) can lead to poor inferences. 

In this paper, we assume that cores are gathered by UIRSing. This makes our results quite general, and covers the wide range of applied cases where UIRSing is used. Furthermore, the variance of sample means from a stratified or cluster sample is typically \textit{lower} than that of a UIRS --- lower variance is the main reason why more sophisticated designs are used. Thus our results can be interpreted as a providing an upper bound on the uncertainty of these other sampling designs. Finally, if we assume that SOC is distributed completely randomly in a given plot (i.e. with no spatial correlation), then the properties of estimates based on a UIRS are equivalent to those based on stratified or cluster random sampling.

There are, however, certain land types or surveys where UIRSing can be logistically infeasible. For example, in row crop studies, only treated rows can be sampled, which is typically much easier to achieve using cluster sampling. Furthermore, note that there is a logistically optimal way to collect $n$ cores by UIRSing. First, sample all $n$ points from the plot, find the shortest path through all $n$ points, and move along that path collecting cores at the sampled points. 
This is called the ``traveling salesman problem" in computer science. The length of the shortest path through a UIRS of size $n$ generated in a plot of area $\mathcal{A}$ tends to be about $0.72 \sqrt{n \mathcal{A}}$ \citep{arlotto_beardwoodhaltonhammersley_2016}. 
Even compared to this shortest path, cluster random samples can have much shorter paths: a transect sample for a rectangular $a \times b$ plot is no longer than $\sqrt{a^2 + b^2}$ for any $n$. 
For example, in the experiments conducted by \citep{tautges_deep_2019} the plots are $64 \times 64$ meters and 10 cores were collected per plot. $\mathcal{A} = 4096$ square meters and the shortest path through $n=10$ randomly generated points is expected to be about 146 meters. On the other hand, a transect through such a plot is about 91 meters. This makes the transect path length only 60\% of the expected length of the best UIRS path.

\section{Compositing}
\label{sec:compositing}

Compositing is the practice of combining cores together from a particular profile in order to capture variability in the plot while reducing assay costs.
Where we call $n$ the number of cores, sampled from the field, $k$ is the number of samples left after compositing.
Edge cases are $n=k$, when we do no compositing, and $k=1$, when we composite down to one sample.
We assume here that each composited sample is comprised of equal proportions of the constituent cores. 
We also assume that $n$ is divisible by $k$ and that each composited sample is comprised of exactly $n/k$ cores.
For example, we might take a UIRS of $n = 30$ cores from a plot and composite down to $k = 6$ composited samples of size $n/k = 5$ constituent cores. 
We also assume that samples are perfectly homogenized after compositing, so that equal parts of constituent samples are present in any given aliquot of the composited sample. Perfect homogenization may be difficult to achieve in some types of soils, like soils with high clay content that tend to clod, which can compromise the validity of compositing.
Our final assumption is compositing additivity, which implies that the SOC concentration in a composited sample is equal to the mean SOC concentration of its constituent cores.
Compositing additivity is met for SOC, but not for other properties like pH, which needs to be considered if investigators plan to measure such properties using the same samples.

There are two reasons why more compositing is not always better.
First, assay error leads to (hopefully) unbiased but still variable assays, which needs to be reduced by assaying multiple cores or else by assaying a single core multiple times. 
Second, compositing is itself an error prone process. It can be very difficult to achieve exactly equal proportions and perfect homogenization, especially in heavy clay soils. 
These challenges can be alleviated and the errors are hedged by assaying more, smaller composite samples.
Finally, in order to do inference we typically need to estimate the plot heterogeneity $\sigma_p$, which can only be estimated when $k \geq 2$, a topic we return to in Section \ref{sec:variance_estimation}.

Logistically, compositing is almost always done in the field to reduce the labor of transporting all $n$ cores to the laboratory. A drawback is that it may be more difficult to achieve good homogenization in field using crude tools on field-moist soil. Furthermore, it is generally important to composite at random. If nearby cores are composited together, which can arise naturally if compositing is done sequentially along a transect or shortest UIRS path, the properties of the sample variance of composited samples may be different. For example, suppose that nearby points tend to have similar SOC concentrations and that nearby points are systematically composited together. In this case the sample variance of composited samples of nearby samples will underestimate $\sigma_p^2$, which will lead to over-optimistic conclusions about the precision of an estimate of $\mu$ \citep{patil_composite_2011}.

\section{Sample Preparation}
\label{sec:sample_prep}

Sample preparation affects both the cost and precision of estimates of $\mu$, and generally depends on the assay method (see Table \ref{tab:sample_prep}).
For dry combustion in an elemental analyzer (DC-EA), samples must be air dried at room temperature. For loss-on-ignition (LOI), samples should be dried in an oven at 105 degrees Celsius, as they must completely dry. The composition of the soil can also determine the proper drying temperature. Salts present in some soils will hold onto water at temperatures higher than 105 degrees, so  \cite{chatterjee_evaluation_2009}.

After drying, samples are passed through a 2mm sieve, which helps remove large bits of organic material (e.g. large roots) and rock. Nevertheless, it can be challenging to differentiate between aggregates and rocks, and to make sure that all $>2$mm aggregate material makes it through the sieve. In particular, some soils are too hard once they dry and must be broken up with a mortar and pestle before they can be sieved.  
Roots may also be picked out by hand. Some studies aim to isolate and separately quantify root fractions. Furthermore, when comparing plots (e.g. in an experiment), carbon in roots can overshadow differences in SOC content \citep{ryals_impacts_2014, fao_measuring_2020}.

After drying, samples are ground to a fine powder (e.g. in a ball mill), which helps ensure homogenization and accurate assay. MIRS can be very sensitive to the size and uniformity of the grind \citep{england_proximal_2018}. On the other hand, LOI does not require soils to be ground.

Finally, many elemental analyzers (EAs) used for DC-EA cannot distinguish between SOC and soil inorganic carbon (e.g. carbonates). For such machines, assays give the concentration of \textit{total} carbon, not just organic carbon. Soils must be checked in advance for inorganic carbon before assay. If the pH is greater than 7.4, ground samples may be treated with hydrochloric acid to remove carbonates \citep{nayak_current_2019}. Methods like LOI don't get hot enough to combust carbonates, while MIRS can usually distinguish between organic and inorganic carbon in spectra.

\section{Assay}
\label{sec:assay}

 In this section, we review the three major methods for assaying SOC concentration before introducing the concept of assay error. For more details on these assay methods, as well as newer \textit{in situ} methods see the recent reviews by \citet{nayak_current_2019} and \citet{england_proximal_2018}.

DC-EA is the gold standard for SOC assay. EAs are expensive to purchase, maintain, and run, but they measure carbon directly and at a fairly high throughput. EAs combust aliquots of soil at high temperatures (around 1000$^{\circ}$ C) in a pure oxygen environment, and assay the CO$_2$ released using gas chromatography. DC-EA is generally the most precise and expensive assay method for SOC, but the precision and cost per analysis will vary by EA model. 

To assay a sample by LOI, investigators measure the mass of dried soil samples, bake them at around 550$^{\circ}$ celsius in a muffle furnace, and then measure how much mass was lost during baking \citep{chatterjee_evaluation_2009, fao_measuring_2020}. This process (ideally) cooks off all the organic matter in the soil, some fraction of which is SOC. The fraction of organic matter that is SOC is determined by calibrating the LOI assays to DC-EA assays using linear regression, or by using a fixed conversion factor of 0.58 \citep{chatterjee_evaluation_2009}. 
However, the nature of the relationship between LOI and DC-EA is often site specific, depending in particular on the vegetation, texture, and residual water content in the soil \citep{vos_capability_2005, nayak_current_2019}.
The site level differences make LOI especially tricky for comparing different plots, as opposed to the same plot at different times, because water content and mineralogy may differ substantially. 
This makes 0.58 suspect as a universally valid fraction.
It is well-known that LOI is relatively imprecise, even in the ideal scenario where it is calibrated to soils using DC-EA. 
However, LOI is considerably cheaper than DC-EA both in terms of upfront costs and costs per sample, and allows investigators to assay many more samples per assay rep than DC-EA \citep{vos_capability_2005}.

MIRS works by shining light in the mid-infrared range (4,000-400 cm$^{-1}$ or 2500-25,000 nm) on dried samples and measuring the wavelengths that are absorbed \citep{nayak_current_2019, reeves_near-_2010, wijewardane_predicting_2018, bellon-maurel_near-infrared_2011}. MIRS is a high-throughput technology that requires even less resources than LOI. It has the further logistical advantage of simultaneously assaying SOC and soil inorganic carbon (SIC), alongside many other soil properties like pH, texture, and cation exchange capacity \citep{wijewardane_predicting_2018}. MIRS is thus a promising new assay method despite the considerable upfront costs of units. Similar to LOI, MIRS must be initially calibrated to DC-EA assays. A database of samples that contains both spectra and DC-EA SOC assays is called a spectral library. Spectra are unique to soils, so spectral libraries must be constructed within a region of interest and do not transfer well to new regions \citep{wijewardane_predicting_2018}. Furthermore, unlike LOI, the relationship between spectra and SOC content is not simple, necessitating the use of more complex prediction methods that need to be rigorously validated \citep{bellon-maurel_near-infrared_2011, wijewardane_predicting_2018}. Calibrations are also highly sensitive to sample prep procedures: samples must be well dried and ground to a consistent size for precise assay \citep{wijewardane_predicting_2018}. Labs can expect to pay a significant upfront cost for purchasing a MIRS unit and establishing a spectral library, but after the initial investment MIRS is cost effective to run, and can be quite precise with proper user training and sample preparation, making it an appealing alternative to DC-EA.

\begin{table}[h]
    \centering
    \begin{tabular}{l | c | c | c}
         Procedure &  LOI & DC-EA & MIRS\\
         \hline
         Transportation & \cmark & \cmark & \cmark \\
         Oven Drying & \cmark & \xmark & \xmark \\
         Air Drying & \xmark & \cmark & \cmark \\
         Sieving & \cmark & \cmark & \cmark \\
         Grinding & \xmark & \cmark & \cmark\\
         Check for IC & \xmark & ~\cmark$^*$ & \xmark
    \end{tabular}
    \caption{A table of sample preparation procedures, their costs per sample, and whether they are needed for assay with LOI, DC-EA, or MIRS. Asterisk denotes that sample preparation may vary depending on specific details of the assay technology or soils. IC = inorganic carbon; LOI = loss-on-ignition; DC-EA = dry combustion in an elemental analyzer; MIRS = mid-infrared spectroscopy.}
    \label{tab:sample_prep}
\end{table}

From a statistical perspective, the assay process is important because additional random error is introduced into the data. Unbiased assays are centered on the true SOC concentration of the (composited) sample. Biased assays systematically overestimate or underestimate the SOC concentration. It is not guaranteed that assays are unbiased (see \citet{bellon-maurel_near-infrared_2011}), though we will assume that they are here. Even when assays are unbiased, they add error to SOC estimation as measurements will not be exactly the same for two or more assays run on the same sample. This variability can be due to errors in weighing, slight differences in aliquots taken from the same sample (especially if homogenization is poor), instrumental drift, or error in predictions or calibrations (especially for LOI and MIRS). We conceptualize assay error on a multiplicative scale so that the amount of error is proportional to the true SOC concentration. Unbiased multiplicative errors are centered at 1, but realizations vary around 1 depending on a variance $\sigma^2_\delta$, which is roughly the expected \textit{percent} error in assay. We detail how to estimate $\sigma_\delta^2$ in Sections \ref{sec:replicate_assays} and \ref{sec:prediction_assay_variance} of the appendix.

As an example, a realized assay error of 1.1 will cause a true SOC concentration of 1\% to appear as 1.1\% and a true SOC concentration of 5\% to appear as 5.5\%. A precise assay method has a small $\sigma_\delta^2$ so realizations tend to be close to 1, and the measured SOC concentration is close to the true SOC concentration. Note that we will sometimes use an additional subscript to refer to a specific method, e.g. $\sigma_{\delta, \tiny{\mbox{DC-EA}}}$ is the assay error variance of DC-EA.

\section{Optimal Sampling and Assay}
\label{sec:optimization}

In this section we highlight our main results. We provide a formula for the precision of estimates of $\mu$ given a sample size $n$ and a number of assays $k$. We derive the optimal $n$ and $k$ that will maximize precision while maintaining a given budget.

\subsection{Estimation Error}
\label{sec:estimation_error}

Suppose we have a UIRS of size $n$ and that composites are formed randomly from $n/k$ samples in equal proportions and with perfect homogenization, so that $k$ assays are taken. Suppose $S_i^*$ is the assayed SOC concentration of the $i$th composited sample. Our estimator is the mean of these assayed composite samples:

$$\hat{\mu} = \frac{1}{k} \sum_{i=1}^k S_i^*.$$
This is an unbiased estimator of $\mu$, so that $\mathbb{E}[\hat{\mu}] = \mu$. Its variance is 

\begin{equation}
    \mathbb{V}(\hat{\mu})= \frac{\sigma_p^2 (1 + \sigma_\delta^2) }{n}  + \frac{\mu^2 \sigma_\delta^2}{k}. \label{eqn:variance_formula}
\end{equation}

If there is no assay error, this reduces to the usual formula for the variance of a UIRS mean: $\sigma_p / n$. Because the estimator is unbiased, it's expected error (mean-squared error) is also equal to (\ref{eqn:variance_formula}). In order to reduce the error, we can either gather more samples $n$ or make more assays $k$. The optimal allocation of samples and assays will depend on the plot parameters $\sigma_p$ and $\mu$, the assay error variance $\sigma_\delta^2$, and a cost model for sampling and assay.

\subsection{Optima}

We now introduce such a cost model. Call $\mbox{cost}_{c}$ the cost of sampling a single core, $\mbox{cost}_P$ the cost of sample preparation, and $\mbox{cost}_A$ the cost of assaying a single composited sample. Note that these costs depend on the sampling and assay methods employed. For example, $\mbox{cost}_A$ under LOI is considerably lower than $\mbox{cost}_A$ under DC-EA. We assume that the cost of compositing itself is negligible, but it could easily be included in $\mbox{cost}_c$. Finally, we assume a fixed cost of the study $\mbox{cost}_0$, which doesn't vary over $n$ and $k$. The total cost is: 

\begin{equation}
   \mbox{cost}_0 + n \cdot \mbox{cost}_{c} + k \cdot (\mbox{cost}_{P} + \mbox{cost}_A).
\end{equation}

We ultimately want to choose both an optimal $n$ and $k$, which we call $n_{\mbox{\tiny opt}}$ and $k_{\mbox{\tiny opt}}$ respectively, as well as a sample prep and assay method. 
We first consider the sample prep and assay methods to be fixed, optimizing only for $n$ and $k$, and then discuss how to choose among strategies. 

Given the cost model along with the plot and assay parameters, the composite size that minimizes the error in Equation (\ref{eqn:variance_formula}) is:

\begin{equation}
\frac{n_{\mbox{\tiny opt}}}{k_{\mbox{\tiny opt}}} = \frac{\sigma_p \sqrt{1 + \sigma_\delta^2}}{\mu \sigma_\delta} \times \sqrt{\frac{\mbox{cost}_P + \mbox{cost}_A}{\mbox{cost}_c}}. \label{eqn:optimal_composite_size}
\end{equation}

The optimal composite size thus depends on the ratio of plot heterogeneity $\sigma_p$ and the degree of assay error $\sigma_\delta$. It also depends on the ratio of assay and sampling costs, though it is less sensitive to small changes in cost due to the square root applied to this ratio. 
Note that there are two boundary conditions that are not reflected in Equation (\ref{eqn:optimal_composite_size}).
Namely, if we initially find $k_{\mbox{\tiny opt}} < 1$ then we take $k_{\mbox{\tiny opt}} = 1$ with the implication that all cores should be fully composited to 1 composite sample. 
On the other hand, if we find $k_{\mbox{\tiny opt}} > n_{\mbox{\tiny opt}}$, then set $k_{\mbox{\tiny opt}} = n_{\mbox{\tiny opt}}$ with the implication that all sampled cores should be assayed without compositing. 
Ultimately, there are only gains to compositing if
$$ \sigma_p^2 (1 + \sigma_\delta^2) (\mbox{cost}_P + \mbox{cost}_A) > \mu^2 \sigma_\delta^2 \mbox{cost}_c.$$
Otherwise no compositing should be done. 

Given a fixed budget $B$, we can compute the optimal variance $\mathbb{V}(\hat{\mu})_{\mbox{\tiny opt}}$. The optimal variance can be difficult to interpret. Taking the square root yields the optimal \textit{standard error} $\mbox{SE}(\hat{\mu})_{\mbox{\tiny opt}}$ we can achieve at budget $B$:
\begin{equation}
    \mbox{SE}(\hat{\mu})_{\mbox{\tiny opt}} =  \frac{\sigma_p \sqrt{(1 + \sigma_\delta^2) \mbox{cost}_c} + \mu \sigma_\delta \sqrt{\mbox{cost}_P + \mbox{cost}_A}}{\sqrt{B - \mbox{cost}_0}} \label{eqn:min_MSE}
\end{equation}
The optimal standard error is on the same scale as the estimate (i.e. percent SOC).

Finally, different sample prep and assay methods involve trade-offs between the costs and the assay error. Clearly, if a method is both cheaper and less erroneous, it is preferred. But how much error should we tolerate for a cheaper assay? The \textit{relative efficiency} of different methods is the ratio of the minimum errors they are able to achieve, per Equation (\ref{eqn:min_MSE}).
The relative efficiency of method 1 over method 2 is:
\begin{equation}
    \frac{\mbox{SE}(\hat{\mu})_{\mbox{\tiny opt}, 1}}{\mbox{SE}(\hat{\mu})_{\mbox{\tiny opt}, 2}} =  \frac{\sigma_p \sqrt{(1 + \sigma_{\delta_1}^2) \mbox{cost}_c} + \mu \sigma_{\delta_1} \sqrt{\mbox{cost}_{P_1} + \mbox{cost}_{M_1}}}{\sigma_p \sqrt{(1 + \sigma_{\delta_2}^2) \mbox{cost}_c} + \mu \sigma_{\delta_2} \sqrt{\mbox{cost}_{P_2} + \mbox{cost}_{M_2}}} \label{eqn:relative_efficiency}
\end{equation}
A relative efficiency close to 1 suggests a near toss-up between different sample prep and assay strategies. On the other hand, a large relative efficiency suggests that method 2 is more efficient than method 1, and vice versa for a small relative efficiency. The upshot is that for any budget, we can achieve substantially more precise estimates when the relative efficiency is far from 1.

Alternatively, given a maximum variance $V$ that we can tolerate, we might ask for a minimum budget over all ways of allocating the budget to samples and assays. This is the inverse of the previous problem. The expressions for the optimum $n$ and $k$ are fairly complicated. We provide details in Section \ref{sec:appendix_minimum_cost} in our appendix.

\subsection{Variance estimation}

\label{sec:variance_estimation}
So far we have assumed that we know the parameters $\sigma_\delta$ and $\sigma_p$. In practice, these quantities must be estimated with gathered data or, when planning a survey, based on physical reasoning and past studies. 

An unbiased estimator of the plot variance $\sigma_p^2$ is the usual sample variance with an adjustment factor for the size of composites:
$$\hat{\sigma}_p^2 = \frac{n}{k} \left [ \frac{1}{k-1}  \sum_{i=1}^k (S_i^* - \hat{\mu})^2 \right ],$$
where as above, $\hat{\mu} = \frac{1}{k} \sum_{i=1}^k S_i^*$
As previously noted, this formula will underestimate the sample variance if composite samples are systematically more homogeneous than the plot itself. This can happen, for example, when composites are grouped together by distance instead of randomly.

We can estimate $\sigma_\delta$ using replicated assays, detailed in Section \ref{sec:replicate_assays} of our appendix. For methods like LOI or MIRS that involve calibration, the additional error due to calibration must be taken into account. See Section \ref{sec:prediction_assay_variance}. 

Putting these pieces together, we can estimate the overall standard error of $\hat{\mu}$ by:

$$\widehat{\mbox{SE}}(\hat{\mu}) = \sqrt{\frac{\hat{\sigma}_p^2 (1 + \hat{\sigma}_\delta^2) }{n}  + \frac{\hat{\mu}^2 \hat{\sigma}_\delta^2}{k}}$$

\subsection{A Confidence Interval}

If the sample size $n$ is not too small, then an asymptotic confidence interval based on the the $t$-distribution with $n-1$ degrees of freedom will be approximately correct. Specifically, denote $t_{(1-\alpha/2)}$ as the $(1-\alpha/2)$ quantile of the $t$-distribution with $n-1$ degrees of freedom. The interval 
$$\left [\hat{\mu} - t_{(1-\alpha/2)} \times \widehat{\mbox{SE}}(\hat{\mu}),~ \hat{\mu} + t_{(1-\alpha/2)} \times \widehat{\mbox{SE}}(\hat{\mu}) \right ]$$ 
bounds the true mean $\mu$ with probability about $(1-\alpha)$. Checking if a particular value of $\mu$ (say $\mu_0$) is in this interval is equivalent to a level $\alpha$ $t$-test of the null hypothesis $H_0: \mu = \mu_0$. Often, a researcher will set $\alpha = .05$ to yield a 95\% confidence interval: $[\hat{\mu} - 1.96 \times \widehat{\mbox{SE}}(\hat{\mu}),~ \hat{\mu} + 1.96 \times \widehat{\mbox{SE}}(\hat{\mu})]$. In instances where the confidence interval includes values less than 0, in particular if $1.96 \times \widehat{\mbox{SE}}(\hat{\mu}) > \hat{\mu}$, it is valid to set the lower confidence limit equal to 0. 

\subsection{Estimating a difference}

Often, investigators aim to estimate the difference between average SOC concentrations, either between two plots at the same time or within the same plot at different times. Let $\mu_1$ and $\mu_2$ be the mean SOC concentrations in plot 1 and plot 2. Then the parameter of interest is $\mu_1 - \mu_2$. Let $\hat{\mu}_1$ and $\hat{\mu}_2$ be estimators of $\mu_1$ and $\mu_2$, as above. Then the difference in means, $\hat{\Delta}_{1,2} = \hat{\mu}_1 - \hat{\mu}_2$, is an unbiased estimator of $\mu_1 - \mu_2$. Furthermore, assuming independent UIRSing in each plot, the standard error is:
$$\mbox{SE}(\hat{\Delta}_{1,2}) = \sqrt{\mathbb{V}( \hat{\mu}_1 ) + \mathbb{V}( \hat{\mu}_2)}.$$ 
The optimum SE of the difference can be attained by separately optimizing $\mathbb{V}( \hat{\mu}_1 )$ and $\mathbb{V}( \hat{\mu}_1 )$, as above, yielding sampling and assay sizes of $n_1$, $k_1$ for plot 1 and $n_2$, $k_2$ for plot 2. A reasonable estimate of the SE is $\widehat{\mbox{SE}}(\hat{\Delta}_{1,2}) \equiv \sqrt{\widehat{\mathbb{V}}( \hat{\mu}_1 ) + \widehat{\mathbb{V}}( \hat{\mu}_2)}$. An approximate $(1-\alpha)$ confidence interval on the difference is: 
$$\left [\hat{\Delta}_{1,2} - t_{(1-\alpha/2)} \times \widehat{\mbox{SE}}(\hat{\Delta}_{1,2}),~ \hat{\Delta}_{1,2} + t_{(1-\alpha/2)} \times \widehat{\mbox{SE}}(\hat{\Delta}_{1,2}) \right ]$$ 
where $t_{(1-\alpha/2)}$ is now the $(1-\alpha/2)$ quantile of the $t$-distribution with $\min(n_1, n_2)$ degrees of freedom.

If sample sizes are fairly small, say $n_1, n_2 < 30$, the difference-in-means will generally not have a normal distribution. In this case, a permutation test should be used to test for a difference between $\mu_1$ and $\mu_2$. Permutation tests provide an exact level $\alpha$ test at any sample size, without assumptions about the distributions of the samples. Permutation confidence intervals can be derived by testing a range of hypotheses over a grid of effect sizes. The corresponding $1-\alpha$ confidence interval contains all effect sizes that are \textit{not} rejected at level $\alpha$. For a review of permutation testing see \citet{good_permutation_2005}.

\section{Application}
\label{sec:application}

In this section we demonstrate a practical application of our analysis. We draw on a variety of sources to estimate parameters and costs. We stress that the results are not intended to provide universal guidance on sampling, sample prep, and assay---they are highly sensitive to the inputs. The open-source software and web tool we provide are intended to enable investigators to draw their own conclusions from their own inputs. 

\subsection{Data}
We combine data from multiple sources to estimate $\sigma_\delta$ for DC-EA, LOI, and MIRS: $\sigma_{\delta, \mbox{\tiny DC-EA}}$,  $\sigma_{\delta, \mbox{\tiny LOI}}$, and  $\sigma_{\delta, \mbox{\tiny MIRS}}$, respectively. $\sigma_{\delta, \mbox{\tiny DC-EA}}$ is estimated from assays on samples taken from rangeland soils in Marin County, California by the Silver Lab at UC Berkeley, referred to here as the Marin data. The samples were run in duplicate on a Carlo Elantech Elemental analyzer at UC Berkeley. We use the method presented in Section \ref{sec:variance_estimation} to compute $\hat{\sigma}_{\delta, \mbox{\tiny DC-EA},i}$ on each sample and took the median across samples to get $\hat{\sigma}_{\delta, \mbox{\tiny DC-EA}}$. We applied the methods presented in detail in Section \ref{sec:prediction_assay_variance} of our appendix to estimate the additional assay error in LOI and MIRS callibrated to DC-EA assays. Briefly, we derived the validation root mean squared error (RMSE$_v$) for LOI by regressing LOI assays on DC-EA assays taken at the Agricultural Diagnostic Laboratory at the University of Arkansas (ADL). These assays were taken on samples from several sites in Colorado collected by the Wainwright Lab at Lawrence Berkeley National Laboratory. We estimated $\sigma_{\delta, \mbox{\tiny MIRS}}$ using the RMSE$_v$ provided in table 4 of \citet{england_proximal_2018}. They computed this estimate from a median of MIRS RMSE$_v$ values reported in a range of studies. These errors were then divided by our estimates of $\hat{\mu}$ and added to the DC-EA error variance estimate to approximate their overall assay error variance on a multiplicative scale. We also computed the SE assuming no assay error and no cost to assay, which represents a typical power analysis and provides a lower bound on the SE across assay methods.

We used the Marin data to get estimates of $\sigma_p$ and $\mu$ in the topsoil (0-10 cm) and in deep soil (50-100 cm). Within depth profiles, we computed the sample mean and standard deviation at each site and then took the median over sites as our estimates $\hat{\sigma}_p$ and $\hat{\mu}$. The samples were collected using transect sampling, not UIRSing, but should provide reasonable estimates.

\subsection{Results}

\subsubsection{Inputs}
All inputs are summarized in table \ref{tab:inputs}. Using the Marin data, we estimated the topsoil plot heterogeneity as $\hat{\sigma}_p = 0.54$ and the mean as $\hat{\mu} = 3.61$. We estimated the deep soil heterogeneity as $\hat{\sigma}_p = 0.12$ and the deep soil mean as $\hat{\mu} = 0.48$. Based on the duplicated DC-EA assays, we obtained the estimate $\hat{\sigma}_{\delta, \mbox{\tiny DC-EA}} = 0.02$.  The RMSE$_v$ for LOI was 0.31 in the range of the Marin data assays. Dividing by $\hat{\mu}$ and combining this with the DC-EA error, we estimated an error variance for LOI of $\hat{\sigma}_{\delta, \mbox{\tiny LOI}} = 0.11$ in the top soil and $\hat{\sigma}_{\delta, \mbox{\tiny LOI}} = 0.67$ in deep soil. \citet{england_proximal_2018} reported a median RMSE$_v$ of 0.11 for MIRS, yielding an estimate of $\hat{\sigma}_{\delta, \mbox{\tiny MIRS}} = 0.05$ in the top soil and $\hat{\sigma}_{\delta, \mbox{\tiny MIRS}} = 0.25$ in deep soil. 

The cost of sampling and the fixed costs of the survey are not well constrained. We set the fixed cost at $\mbox{cost}_0 \equiv 200$ and $\mbox{cost}_c$ at 5, 20, or 40 USD to reflect cheap, medium, and expensive sampling. We assumed a transport cost of 2.00, a cost of 4.00 for oven drying, 1.00 for air drying, 2.00 for sieving, 4.00 for grinding, and 2.00 for acid testing for inorganic carbon. Without root picking, this puts the cost of sample prep at 8.00 for LOI, 11.00 for DC-EA, and 9.00 for MIRS. The assay costs for DC-EA and MIRS were estimated in \citep{rourke_optical_2011}. That paper reported a cost of about 15 USD per sample for DC-EA and 1.30 USD per sample for MIRS. The 12:1 price ratio of DC-EA to LOI reported in \citet{vos_capability_2005} yields an assay cost of 1.25 USD per sample for LOI. Adding $\cost_P$ and $\cost_A$ for each method yields $\mbox{cost}_{\tiny \mbox{DC-EA}} = 26.00$ USD, $\mbox{cost}_{\tiny \mbox{LOI}} = 9.25$ USD, and $\mbox{cost}_{\tiny \mbox{MIRS}} = 10.30$ USD. 

\begin{table}[]
    \centering
    \begin{tabular}{l|l|l}
        \textbf{Description} & \textbf{Notation} & \textbf{Value(s)}\\
        \hline
        DC-EA assay variance & $\hat{\sigma}_{\delta, \mbox{\tiny DC-EA}}$ & 0.02 \\
        LOI assay variance & $\hat{\sigma}_{\delta, \mbox{\tiny LOI}}$ & 0.11, 0.67\\
        MIRS assay variance &$\hat{\sigma}_{\delta, \mbox{\tiny MIRS}}$ & 0.05, 0.25\\  
        Plot heterogeneity & $\hat{\sigma}_p$ & 0.68, 0.12\\
        Plot mean & $\hat{\mu}$ & 3.57, 0.48\\
        \hline
        Fixed cost & $\mbox{cost}_0$ & 200\\
        Cost of DC-EA assay & $\mbox{cost}_{\tiny \mbox{DC-EA}}$ & 26.00 \\
        Cost of LOI assay & $\mbox{cost}_{\tiny \mbox{LOI}}$ & 9.25 \\
        Cost of MIRS assay & $\mbox{cost}_{\tiny \mbox{MIRS}}$ & 10.30\\
        Cost of taking a core & $\mbox{cost}_c$ & 5, 20, 40 \\
    \end{tabular}
    \caption{Inputs to optimization problem as estimated from Marin and LBL data. $\hat{\sigma}_{\delta, \mbox{\tiny DC-EA}}$ is the assay error of DC-EA; $\hat{\sigma}_{\delta, \mbox{\tiny LOI}}$ is the assay error of LOI; $\hat{\sigma}_p$ is the plot heterogeneity (standard deviation); $\hat{\mu}$ is the plot mean concentration; $\mbox{cost}_{\tiny \mbox{DC-EA}}$ is the cost of DC-EA assay plus the costs of associated sample prep; $\mbox{cost}_{\tiny \mbox{LOI}}$ is the cost of LOI plus the costs of associated sample prep;  $\mbox{cost}_{\tiny \mbox{MIRS}}$ is the cost of MIRS plus the costs of associated sample prep. $\mbox{cost}_c$ is the cost of sampling.}
    \label{tab:inputs}
\end{table}

\subsubsection{Outputs}
\begin{figure}[!p]
    \centering
    \includegraphics[width = \textwidth]{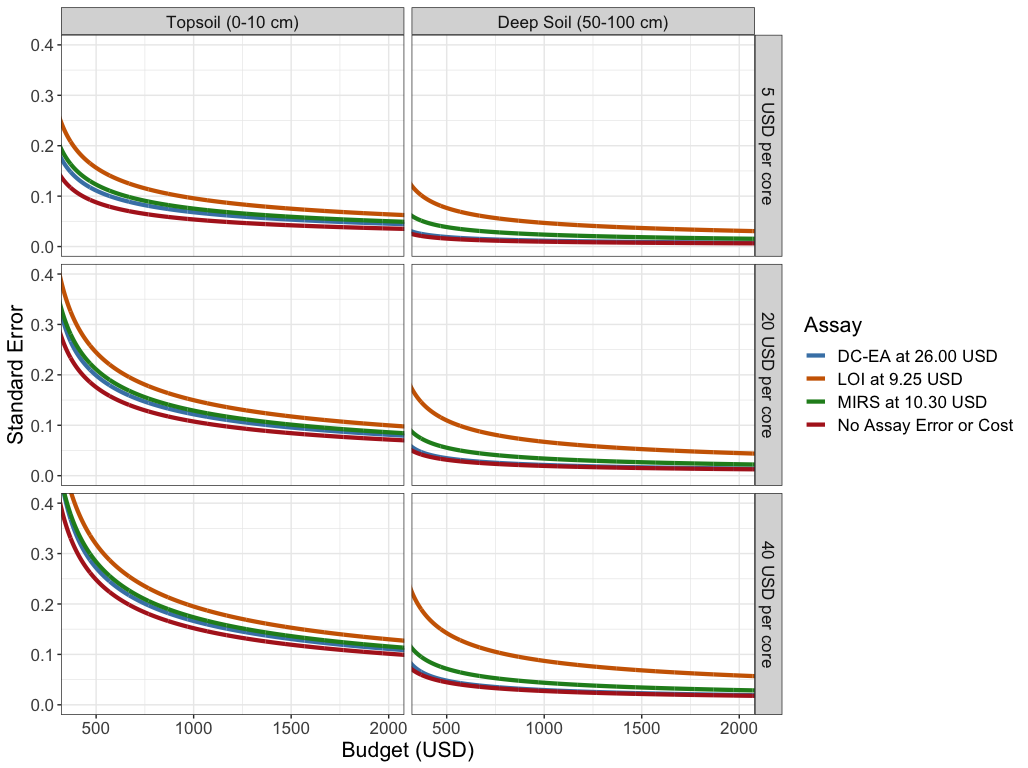}
    \caption{Optimal standard errors for estimating $\mu$ given parameters in Table \ref{tab:inputs}. The x-axis is the budget in US dollars, the y-axis is the standard error in \%SOC attained at the given budget. Different colored lines correspond to different assay methods. The cost of sampling varies across rows, and the depth varies across columns. The line labels indicate the combined costs of sample prep and assay for each method. DC-EA = dry combustion in an elemental analyzer; LOI = loss-on-ignition; MIRS = mid-infrared spectroscopy; USD = United States dollars.}
    \label{fig:optima_variance_plot}
\end{figure}

\begin{figure}[!p]
    \centering
    \includegraphics[width = \textwidth]{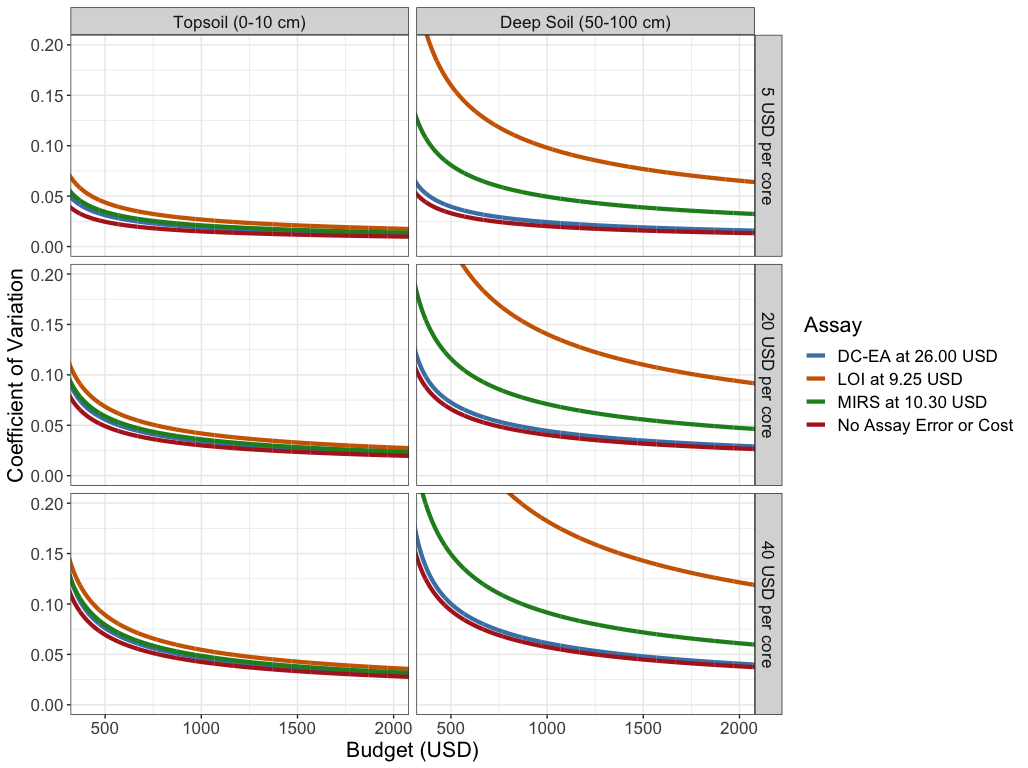}
    \caption{Optimal coefficients of variation for estimating $\mu$ given parameters in Table \ref{tab:inputs}. The x-axis is the budget in US dollars, the y-axis is the coefficient of variation: $\mbox{SE}(\hat{\mu})_{\mbox{\tiny opt}} / \mu$. Different colored lines correspond to different assay methods. The cost of sampling varies across rows, and the depth varies across columns. The line labels indicate the combined costs of sample prep and assay for each method. DC-EA = dry combustion in an elemental analyzer; LOI = loss-on-ignition; MIRS = mid-infrared spectroscopy; USD = United States dollars.}
    \label{fig:optima_cv_plot}
\end{figure}

 Figure \ref{fig:optima_variance_plot} plots the optimal SE of estimation attainable for each assay method across a range of budgets. Figure \ref{fig:optima_cv_plot} plots the same results but rescaling SEs to coefficients of variation. The output indicates that DC-EA is the best assay method in both topsoil and deep soil, yielding the most precise estimate at any given budget. In terms of relative performance, the assay method is more important in the deep soil than in the topsoil: DC-EA represents a major improvement over the other methods in deep soil, while the precision is essentially a toss-up in top soil. Under our inputs, DC-EA gets close to achieving the lower bound implied by no assay error. 

Optimal composite sizes are provided in Table \ref{tab:optimal_composites} across the range of sampling costs and depths. Compositing is more valuable as the assay method becomes more precise and expensive, with large gains to compositing under DC-EA and essentially no gain under LOI. Compositing is also more valuable if samples are cheap to gather and the plot is heterogeneous, in which case it becomes beneficial to focus budgets on sampling rather than assay.     

Relative efficiencies are given in Table \ref{tab:relative_efficiency}. Relative efficiencies for topsoil assays are fairly close to 1. For deep soil, DC-EA is at least twice as efficient as LOI --- all relative efficiencies are less than 0.5 --- and at least 30\% more efficient than MIRS for any sampling cost.

\begin{table}[h]
    \centering
    \begin{tabular}{c|c |c| c|c |c |c}
        & \multicolumn{3}{c}{Topsoil (0-10 cm)} & \multicolumn{3}{|c}{Deep Soil (50-100 cm)}\\
          & 5 USD & 20 USD & 40 USD & 5 USD & 20 USD & 40 USD \\
          \hline
         DC-EA & 20 & 10 & 7 & 26 & 13 & 9\\
         MIRS & 5 & 3 & 2 & 2 & 1 & 1\\
         LOI & 2 & 1 & 1 & 1 & 1 & 1
    \end{tabular}
    \caption{Optimal composite sizes for the three assay methods. Sampling costs are set at 5, 20, or 40 USD (top row). Soil parameters are determined from the topsoil (first three columns) or deep soil (last three columns). An optimal composite size of 1 suggests that no compositing should be done, i.e. that all cores should be measured. }
    \label{tab:optimal_composites}
\end{table}

\begin{table}[ht]
\centering
\begin{tabular}{rrcc}
\textbf{Soil Profile} & \textbf{Sampling Cost} & \textbf{SE$_{\mbox{\tiny DC-EA}}$ / SE$_{\mbox{\tiny LOI}}$} & \textbf{SE$_{\mbox{\tiny DC-EA}}$ / SE$_{\mbox{\tiny MIRS}}$} \\ 
  \hline
Topsoil (0-10 cm) & 5 USD & 0.70 & 0.90 \\ 
  Topsoil (0-10 cm) & 20 USD & 0.79 & 0.93 \\ 
  Topsoil (0-10 cm) & 40 USD & 0.83 & 0.95 \\ 
  \hline
  Deep Soil (50-100 cm)& 5 USD & 0.25 & 0.48 \\ 
  Deep Soil (50-100 cm)& 20 USD & 0.36 & 0.61 \\ 
  Deep Soil (50-100 cm)& 40 USD & 0.42 & 0.67 \\  
   \hline
\end{tabular}
\label{tab:relative_efficiency}
\caption{Relative efficiencies of different assay methods compared to DC-EA at different profiles (first column) and sampling costs (second column). A relative efficiency significantly less than 1 suggests DC-EA is more efficient than the alternative method at any given budget, and vice versa for a relative efficiency greater than 1. A relative efficiency near 1 suggests little difference between methods.}
\end{table}

\section{Discussion}
\label{sec:discussion}

In this paper, we statistically formalized the sampling and assay processes to characterize the precision of SOC concentration estimation while making minimal assumptions. We derived optimal composite sizes to maximize precision under a fixed budget. Although we did not discuss it extensively, we also solved the inverse problem of minimizing costs given a fixed precision (see section \ref{sec:appendix_minimum_cost} of our appendix). A graphical user-interface is available at \url{https://jakespertus.shinyapps.io/shiny/}, and the source code can be found at \url{https://github.com/spertus/soil-carbon-statistics}. 

We applied our method to data from a California rangeland, bringing in costs and errors of measurement from other studies \citep{vos_capability_2005, england_proximal_2018, rourke_optical_2011}. There are a number of interesting implications from our results. 

First, we found assay error to be a significant source of uncertainty in SOC estimation that is usually not taken into account. Indeed, many analyses in the SOC literature compute uncertainty estimates accounting only for plot heterogeneity (or in some cases, only inter-plot heterogeneity). We found that incorporating assay error and costs can double the uncertainty (in terms of standard error) compared to the conventional approach of not incorporating assay error. 

Furthermore, the depth of soil under study was an especially important consideration for the assay method employed. We found that efficiencies varied much more across the assay methods when attempting to quantify deep soil concentrations rather than top soil. In terms of the coefficient of variation, top soil can be quantified by any assay method to within about 5\% of the mean at a budget of 1000 USD. On the other hand, DC-EA seems far better at accurately quantifying deep soil concentrations than other methods despite its high cost. Equation $\ref{eqn:min_MSE}$ reveals that when the plot heterogeneity $\sigma_p$ is high the estimation error will be driven largely by the cost of sampling while the cost and precision of assay have little effect. Intuitively, we need many samples to characterize the heterogeneity within the plot, and cheaper, less precise assay methods generally allow many samples to be collected and assayed. Conversely, if the plot heterogeneity is low, it is better to collect a few samples that accurately represent the average plot concentration and focus the budget on assaying them as precisely as possible.

We also found that with our inputs, the benefits of compositing were quite variable. Compositing many cores together is beneficial when the assay method is fairly expensive and precise (e.g. DC-EA), while sampling is fairly cheap (e.g. 5.00 USD per core). Equation \ref{eqn:optimal_composite_size} reveals that compositing may also be resourceful when the plot heterogeneity is large compared to assay error. 

We did not incorporate bulk density into our analysis. Estimating bulk density is critical to converting from concentrations to stocks. Estimating SOC stock is especially necessary in studies of carbon sequestration and climate change mitigation, while SOC concentration is typically the parameter of interest from a soil health and functioning perspective. As with SOC concentrations, bulk density tends to vary substantially across a landscape. However, investigators frequently take only one bulk density sample and bulk density is also prone to assay error \citep{walter_determining_2016}. Thus converting from concentration to stock will incur substantial additional error, which should ideally be reflected in confidence intervals on stock estimates. Our results on the error in concentration estimation can be seen as a lower bound on the error in stock estimation, i.e. assuming no error in bulk density estimation. In addition to including error in bulk density, future work should incorporate more general sampling schemes that can improve efficiency, like stratified sampling or well-spread sampling, and optimize over the sampling design as well as the assay method. Considering the sampling design alongside the compositing and assay strategy will allow investigators to design economical soil surveys that achieve their desired precision.  

\section*{Acknowledgements}

I thank Philip Stark for his advice and mentorship. Whendee Silver and Haruko Wainwright provided their data and perspectives. This paper would not be possible without numerous conversations with Paige Stanley and Jessica Chiartas, who also provided enormously helpful feedback on an earlier draft of this paper. 

\section*{Funding}

Jacob Spertus is supported by a National Science Foundation (NSF) Graduate Research Fellowship (DGE 1752814). 

\bibliography{soilcarbonstatistics}

\appendix

\section{Mathematical Framework}
\label{sec:framework}

We have a plot $\mathcal{P} \subset \mathbb{R}^3$. An element of $\mathcal{P}$ is a 3-tuple $(x,y,z)$. $x$ denotes longitude or $x$-axis distance from an origin (e.g. the lower left hand corner of a rectangular plot), $y$ denotes latitude or $y$-axis distance, and $z$ denotes depth. 

At $(x,y,z)$ the soil has some concentration of SOC, which we will denote by $c(x,y,z) \in [0,100]$ with units in percent or equivalently grams SOC per hectogram of soil. Note that sometimes SOC concentration is reported in grams per kilogram. Note also that $c(x,y,z)$ is best conceptualized as an average over a small window centered at $(x,y,z)$. Taking the design-based perspective, we consider $c(x,y,z)$ to be fixed but unknown.

To convert to grams SOC per volume of soil, take $d(x,y,z)$ to be the density of the soil at point $(x,y,z)$, e.g. in grams per cubic centimeter. This is called the ``bulk density" in soil science. The amount or \textit{stock} of carbon in a small area centered at point $(x,y,z)$ is thus $c(x,y,z) \times d(x,y,z)$. The total amount or stock of carbon in a plot is:

$$\mathcal{T} = \int_\mathcal{P} c(x,y,z) \times d(x,y,z) d\mathcal{P}  = \int_0^{x_{\max}} \int_0^{y_{\max}} \int_0^{z_{\max}} c(x,y,z) \times d(x,y,z) ~dz ~ dy ~dx  $$
Assuming constant bulk density means that $d(x,y,z) = d$ and the total carbon becomes:
$$\mathcal{T} = d \int_\mathcal{P} c(x,y,z) d\mathcal{P} = d \times \mu$$ 
where $\mu \equiv \int_\mathcal{P} c(x,y,z) d\mathcal{P}$ is the population average SOC concentration---the key parameter to be estimated through soil sampling. The bulk density $d$ must also be estimated.

The population variance of a plot is formally:
$$\sigma_p^2 = \int_\mathcal{P} \left [c(x,y,z) - \mu \right]^2 d\mathcal{P}$$
The population variance is a measure of heterogeneity that is instrumental in determining the precision of estimates of $\mu$.

$\mu$ and $\sigma_p$ are estimated using sampled cores. The plot is often sliced into profiles along depth, and positions $(x,y)$ locations are randomly sampled within depth. From here on we will assume we are sampling within a profile and ignore depth. Randomly sampled positions are denoted $\{(X_i, Y_i)\}_{i=1}^n$ and the $n$ corresponding cores are denoted $\{c(X_i, Y_i)\}_{i=1}^n$, or $\{C_1,...,C_n\}$ when the location is not important. We suppose here that these cores are selected by UIRSing. The properties of the sample mean of cores from a UIRS, $\bar{C} = \frac{1}{n} \sum_{i=1}^n C_i$, are simple and well-understood: $\bar{C}$ is unbiased ($\mathbb{E}[\bar{C}] = \mu$) and has variance $\mathbb{V}[\bar{C}] = \sigma_p^2 / n$.

Given a UIRS $\{C_1,...,C_n\}$ compositing bins the cores into $k$ groups of size $n/k$. The groups are denoted by a set of indices $\{g_1,...,g_k\}$, where $g_1 = \{1,...,n/k\}$, $g_2 = \{n/k+1,...,2n/k\}$, etc. The cores in each group are physically mixed together to form composite samples $\{S_1, ... S_k\}$. Under compositing additivity, we have $S_i = \frac{k}{n} \sum_{j \in g_i}  C_j$. Note also that the above notation covers the case where no compositing is done, with $k = n$ and $S_i = C_i$.

Under equal proportions compositing of cores from a UIRS, it follows immediately that the sample mean of the composited cores is an unbiased estimate of $\mu$:
$$\mathbb{E}\bigg [\frac{1}{k} \sum_{i=1}^k S_i  \bigg ] = \mathbb{E}\bigg [\frac{1}{k} \sum_{i=1}^k \sum_{j \in g_i} \frac{k}{n} C_j \bigg ] = \mathbb{E}\bigg [\frac{1}{n} \sum_{i=1}^n C_i  \bigg ] = \mu$$
Furthermore, because the sample mean of the composite samples is equivalent to the sample mean of the constituents, it's variance is also 
$$\mathbb{V}\bigg[ \frac{1}{k} \sum_{i=1}^k S_i \bigg ] = \mathbb{V}\bigg [\frac{1}{n} \sum_{i=1}^n C_i \bigg ] = \frac{\sigma_p^2}{n}.$$

Assay error is denoted $\delta_i$. Measured samples are the true average concentrations, perturbed multiplicatively by the assay error: $S_i^* = S_i \delta_i$.
We assume assay errors are independent of true concentrations and unbiased so that $\mathbb{E}(\delta_i) = 1$ and hence $\mathbb{E}(S_i^*) = \mathbb{E}[S_i] = \mu$. We also assume that the assay error has constant variance $\mathbb{V}(\delta_i) = \sigma_\delta^2$, that does not depend on $S_i$. 

Our estimator is the mean of $k$ measured samples composited from $n$ cores:
$\hat{\mu} = \frac{1}{k} \sum_{i=1}^k S_i^*$. Under our assumptions, $\hat{\mu}$ is an unbiased estimator: 
\begin{equation*}
   \mathbb{E}\left [\frac{1}{k} \sum_{i=1}^k S_i\delta_i \right ]
=  \mathbb{E}\left [\frac{1}{k} \sum_{i=1}^k \left [ \sum_{j \in g_i} \frac{k}{n} C_j \right ] \delta_i \right ]
= \frac{1}{k} \sum_{i=1}^k \left [ \sum_{j \in g_i} \frac{k}{n}\mathbb{E}[C_j] \right ] \mathbb{E}[\delta_i]
= \frac{1}{k} \sum_{i=1}^k \left [ \sum_{j \in g_i} \frac{k}{n}\mu \right ] 
= \mu. 
\end{equation*}
Its variance is:
\begin{align*}
\mathbb{V}\left [\hat{\mu} \right ] 
&= \frac{1}{k^2} \sum_{i=1}^k \mathbb{V}[S_i \delta_i]\\
&= \frac{1}{k^2}  \sum_{i=1}^k \left ( \mathbb{V}[S_i] \mathbb{V}[\delta_i] + \mathbb{E}[S_i]^2 \mathbb{V}[\delta_i] + \mathbb{E}[\delta_i]^2 \mathbb{V}[S_i] \right )\\
&= \frac{1}{k^2}  \sum_{i=1}^k \left ( \frac{k}{n} \sigma^2 \sigma_\delta^2 + \mu^2 \sigma_\delta^2 +  \frac{k}{n} \sigma^2\right )\\
&=  \frac{\sigma^2 (1 + \sigma_\delta^2) }{n}  + \frac{\mu^2 \sigma_\delta^2}{k}\\ 
\end{align*}

\section{Optimizations}

\subsection{Minimum error with a fixed budget}

For a budget $B$, fixed in advance, we seek the solution to the optimization problem:

\begin{alignat}{2}
      \mathbb{V}(\hat{\mu})_{\mbox{\tiny opt}} = &\min_{M,P} \min_{n,k}~~ &&\frac{\sigma_p^2 (1 + \sigma_\delta^2) }{n}  + \frac{\mu^2 \sigma_\delta^2}{k} \label{eqn:mse} \\
      &\mbox{st:} &&\mbox{cost}_0 + n \cdot \mbox{cost}_{c} + k \cdot (\mbox{cost}_{P} + \mbox{cost}_A) \leq B\\
      &~ &&k \geq 1 \label{eqn:positivity_constraint}\\
      &~ &&k \leq n \label{eqn:k_less_than_n}
\end{alignat}
where as above $\mbox{cost}_0$ is the fixed cost, $\mbox{cost}_{c}$ is the cost of sampling a single core, $\mbox{cost}_{P}$ is the cost of sample prep, and $\mbox{cost}_{A}$ is the cost of assay. $A$ and $P$ additionally denote the assay and sample preparation schemes, which affect costs and $\sigma_\delta$. 

For a fixed $A$ and $P$, the inner optimization problem can be solved in closed form using a Lagrange multiplier for the constraint. The optimal sampling and assay sizes are then: 

\begin{align}
    n_{\mbox{\tiny opt}} &= ~ (B - \mbox{cost}_0) \frac{ \sigma_p \sqrt{1 + \sigma_\delta^2}}{[\sigma_p \sqrt{ (1+\sigma_\delta^2) \mbox{cost}_c} + \mu \sigma_\delta \sqrt{\mbox{cost}_P + \mbox{cost}_A}] \sqrt{\mbox{cost}_c}}  \label{eqn:n}\\
     k_{\mbox{\tiny opt}} &=  (B - \mbox{cost}_0) \frac{ \mu \sigma_\delta}{[\sigma_p \sqrt{(1+\sigma_\delta^2) \mbox{cost}_c} + \mu \sigma_\delta \sqrt{\mbox{cost}_P + \mbox{cost}_A}] \sqrt{\mbox{cost}_P + \mbox{cost}_A}} \label{eqn:k}.
\end{align}

This solution ignores the constraints $k \geq 1$ and $k \leq n$. If we find $k_{\mbox{\tiny opt}} < 1$, then set $k_{\mbox{\tiny opt}} = 1$ and $n_{\mbox{\tiny opt}} = (B - \mbox{cost}_0 - \mbox{cost}_P - \mbox{cost}_A) / \mbox{cost}_c$. If we find $k_{\mbox{\tiny opt}} > n^*_{P,M}$ then set $k_{\mbox{\tiny opt}} = n_{\mbox{\tiny opt}} = (B - \mbox{cost}_0) / (\mbox{cost}_c + \mbox{cost}_P + \mbox{cost}_A)$. To obtain integer solutions while staying under budget, $n_{\mbox{\tiny opt}}$ and $k_{\mbox{\tiny opt}}$ should be rounded down.


\subsection{Minimum cost for a given precision}
\label{sec:appendix_minimum_cost}

Given a maximum variance $V$ that we can tolerate, we seek the minimum budget over all ways of allocating the budget to samples and assays while achieving that precision. Formally:

\begin{alignat}{2}
      B_{\mbox{\tiny opt}} = &\min_{M,P} \min_{n,k}~~ &&\mbox{cost}_0 + n \cdot \mbox{cost}_{c} + k \cdot (\mbox{cost}_P + \mbox{cost}_A) 
       \\
      &\mbox{st:} &&\frac{\sigma_p^2 (1 + \sigma_\delta^2) }{n}  + \frac{\mu^2 \sigma_\delta^2}{k} \leq V\\
      &~ &&k \geq 1\\
      &~ &&k \leq n
\end{alignat}

The solution of the inner optimization (for fixed $M$ and $P$) is:

\begin{align}
    n_{\mbox{\tiny opt}} &=  \left [ \left(\frac{\mbox{cost}_c \cdot  \sigma_p^2 (1+\sigma_\delta^2)}{\mbox{cost}_P + \mbox{cost}_A} \right)^{1/2} + 1 \right ] \frac{\mu^2 \sigma_\delta^2}{V}  \\
     k_{\mbox{\tiny opt}} &=   \frac{\sigma_p^2 (1+\sigma_\delta^2)}{V \left(1 - \left [ \left(\frac{\mbox{cost}_c \cdot  \sigma_p^2 (1+\sigma_\delta^2)}{\mbox{cost}_P + \mbox{cost}_A} \right)^{1/2} + 1 \right ]^{-1} \right )}.
\end{align}

The constraints $k \geq 1$ and $k \leq n$ are not respected by these solutions. If we find $k_{\mbox{\tiny opt}} < 1$, then set $k_{\mbox{\tiny opt}} = 1$ and $n_{\mbox{\tiny opt}} = \sigma^2_p / V$ (obtained, for example, when there is no assay error). If we find $k_{\mbox{\tiny opt}} \geq n_{\mbox{\tiny opt}}$, then set $k_{\mbox{\tiny opt}} = n_{\mbox{\tiny opt}} = (\sigma_p^2 (1 + \sigma_\delta^2) + \mu^2 \sigma_\delta^2) / V$. To get integer solutions $n_{\mbox{\tiny opt}}$ and $k_{\mbox{\tiny opt}}$ should be rounded up.

\section{Estimating $\sigma_\delta^2$}

\subsection{Replicate assays}
\label{sec:replicate_assays}

Suppose we have $r$ replicated, unbiased assays for the $i$th sample. The replicates are denoted $\{S^*_{i1}, S^*_{i2},..., S^*_{i r}\}$, where $S^*_{i j} = S_i \delta_{ij}$ is the true SOC concentration in composite sample $i$ multiplied by an independent, mean 1 assay error. The sample mean over replicates is $\bar{S}_i = \frac{1}{r} \sum_{j=1}^r S^*_{i j}$, which is an unbiased estimate of $S_i$ with variance $\mathbb{V}[\bar{S}_i | S_i] = \mathbb{V}[S^*_{ij} | S_i] / r$. An unbiased estimate of $S_i^2$ is the squared sample mean minus its variance, i.e. $\bar{S}_i^2 - \frac{1}{r(1-r)} \sum_{j=1}^r (S^*_{ij} - \bar{S}_i)^2$. 
Thus we might estimate $\sigma_\delta^2$ by plugging in the unbiased estimators of $\mathbb{V}[S_{ij}^* | S_i]$ and $S_i^2$:
\begin{equation}
    \hat{\sigma}^2_\delta = \frac{\frac{1}{r-1} \sum_{j=1}^{r} (S_{ij}^* - \bar{S}_i^*)^2}{(\bar{S}_i^*)^2 - \frac{1}{r (r-1)} \sum_{j=1}^{r} (S_{ij}^* - \bar{S}_i^*)^2}. \label{eqn:sigma_delta_hat}
\end{equation}
This is not necessarily unbiased. If the numerator and denominator are independent, then Jensen's inequality shows this is conservative in expectation: $\mathbb{E}[\hat{\sigma}^2_\delta] > \sigma_\delta$. We used this replicated measurement technique to estimate the error of DC-EA.

$\hat{\sigma}^2_{\delta}$ can be computed on any sample that is replicated 2 or more times. One strategy to estimate $\sigma_\delta$ is thus to duplicate every sample ($r=2$) and then take the average or median, though the variance of estimates may be high.
Plotting $\hat{\sigma}^2_{\delta i}$ against $S_i^*$ should indicate potential violations of the constant assay error variance assumption. 
Another strategy is to replicate a single sample some large number of times, say $r = 30$, but this will not provide information about the constant variance assumption. 
A good balance is to measure a few samples along a grid of $S_i^*$ values some moderately large number of times, say $r = 5$. Under constant assay error variance, the estimates $\hat{\sigma}^2_{\delta i}$ should be fairly close and there should not be a trend in $S_i^*$.

\subsection{Prediction Methods}
\label{sec:prediction_assay_variance}
Suppose we have a method that is calibrated to an unbiased assay (e.g. DC-EA) by regression, like LOI or MIRS. There are two sources of error in the estimate. First, there is the assay error of the calibration assay which can be estimated directly through replication as per Section \ref{sec:replicate_assays}. Second, there is the error in the calibration itself, i.e. prediction error. We discuss two ways to estimate the prediction error. To approximate the total error of a prediction method we recommend simply adding the pieces together.

Prediction methods typically assume an additive error model and estimate the variance of the additive error out of sample. We will call this estimate $\mbox{RMSE}_v$ for validation root mean squared error. Now, if SOC concentration has been first transformed to the log scale, then the model implicitly assumes that error is multiplicative on the original scale, our estimate of the error in prediction is then $\exp(\mbox{RMSE}_v)$. On the other hand, if SOC is modeled directly (i.e. without a log transformation) then we can approximate the error on a multiplicative scale by assuming that the additive error pertains to the average SOC assay, which suggests dividing by the average assay. Thus we estimate the prediction error by $\frac{\mbox{RMSE}_v}{\hat{\mu}}$. 

In our application, LOI and MIRS were calibrated both to DC-EA assays. For both of these methods, we had an RMSE$_v$ of SOC modeled on the original scale (without a log transform) so we estimated the error of these methods as 
\begin{align*}
    \hat{\sigma}_{\delta, \mbox{\tiny LOI}} &= \hat{\sigma}_{\delta, \mbox{\tiny DC-EA}} + \frac{\mbox{RMSE}_{v, \mbox{\tiny LOI}}}{\hat{\mu}}\\
    \hat{\sigma}_{\delta, \mbox{\tiny MIRS}} &= \hat{\sigma}_{\delta, \mbox{\tiny DC-EA}} + \frac{\mbox{RMSE}_{v, \mbox{\tiny MIRS}}}{\hat{\mu}}\\
\end{align*}

\section{Shortest path through random points in $\mathcal{P}$}

Recall that the area of plot $\mathcal{P}$ is $\mathcal{A}$. Finding the shortest path through $n$ points is known as the traveling salesman problem. If the $n$ points are generated randomly and independently with density $f(x)$ then the Beardwood-Halton-Hammersley theorem for $\mathbb{R}^2$ says that the length of the shortest path converges to:
$$L_n / \sqrt{n} \rightarrow \beta_2 \int_{\mathbb{R}^2} \sqrt{f(x)} dx$$

$\beta_2$ is an unknown constant, but analytical bounds and numerical simulations have pegged it at about 0.714 \citep{arlotto_beardwoodhaltonhammersley_2016}. Under UIRSing, $f(x) = \frac{1}{\mathcal{A}}$ on $\mathcal{P}$ and 0 elsewhere. The integral thus evaluates to $\sqrt{\mathcal{A}}$. 

\end{document}